\documentclass[pmlr]{jmlr}
\usepackage{gtml2025_workshop}
\jmlrvolume{325}
\jmlryear{2025}
\jmlrworkshop{Geometry, Topology, and Machine Learning}
\editors{Michael Bleher, Freya Jensen, Levin Maier, Diaaeldin Taha, and Anna Wienhard}
\firstpageno{145}
\usepackage[utf8]{inputenc}
\usepackage[T1]{fontenc}
\usepackage{url}
\usepackage{graphicx}
\usepackage{amsmath, amssymb}
\usepackage{booktabs}

\title[Geometry of Interbrain Networks]{On a Geometry of Interbrain Networks}

\author{
\Name{Nicol\'{a}s Hinrichs}
\Email{hinrichsn@cbs.mpg.de}\\
\addr Neural Data Science and Statistical Computing Group, Max Planck Institute for Human Cognitive and Brain Sciences, Leipzig, Germany; Embodied Cognitive Science Unit, Okinawa Institute of Science and Technology, Okinawa, Japan
\AND
\Name{Noah Guzm\'{a}n}
\Email{nguzman313@gmail.com}\\
\addr Independent Scholar
\AND
\Name{Melanie Weber}
\Email{mweber@g.harvard.edu}\\
\addr School of Engineering and Applied Sciences, Harvard University, Cambridge, MA, United States
}

\date{}

\begin{document}

\maketitle

\begin{abstract}
Effective analysis in neuroscience benefits significantly from robust conceptual frameworks. Traditional metrics of interbrain synchrony in social neuroscience typically depend on fixed, correlation-based approaches, restricting their explanatory capacity to descriptive observations. Inspired by the successful integration of geometric insights in network science, we propose leveraging discrete geometry to examine the dynamic reconfigurations in neural interactions during social exchanges. Unlike conventional synchrony approaches, our method interprets inter-brain connectivity changes through the evolving geometric structures of neural networks. This geometric framework is realized through a pipeline that identifies critical transitions in network connectivity using entropy metrics derived from curvature distributions. By doing so, we significantly enhance the capacity of hyperscanning methodologies to uncover underlying neural mechanisms in interactive social behavior.
\end{abstract}


\section{Introduction}
\label{sec:intro}

Interbrain synchrony (IBS) metrics, such as the Phase Locking Value (PLV), have dominated social neuroscience research, providing practical but fundamentally descriptive measures of neural interactions \citep{Hakim2023}. These methods typically neglect dynamic transitions between brain network states that could provide insight into social interaction mechanisms. Recent advancements in geometric machine learning highlight discrete geometric methods as powerful tools for characterizing complex network structures and dynamics~\citep{Weber2025}. The present opinion piece is motivated by the idea that transient connectivity patterns govern flexible cognitive processes~\citep{Sporns2010}. Such processes have been previously analyzed with geometric tools; however, these works explored primarily intra-brain structural and functional networks \citep{chatterjee2021detecting,Weber2019}. In this article, we propose the application of geometric methods to time-varying interbrain networks during social interaction. Specifically, our proposed approach leverages discrete graph curvatures to address the unique challenges of dynamic interbrain networks in hyperscanning research \citep{Hinrichs2025}; it aims to overcome the limitations of correlation-based metrics by providing richer, more mechanistic insights into how brain networks dynamically reorganize during social interactions. 

\section{A Graph Geometry Toolkit}
\label{sec:Geo}

Central to our proposal are discrete curvatures, one example of which is the Forman-Ricci curvature (FRC). Developed initially to characterize geometric properties of discrete spaces parametrized as cell complexes~\citep{Forman2003}, a specialization of FRC to graphs quantifies the expansion and contraction of information across the network by examining the network’s connectivity patterns. Specifically, the FRC of an edge \( e \) connecting nodes \( i \) and \( j \) in a weighted network is defined as
\begin{equation}
    F(e) = w_e\left(\frac{z_i}{w_e} + \frac{z_j}{w_e} - \sum_{e_i \sim i, e_i \neq e}\frac{z_i}{\sqrt{w_e w_{e_i}}} - \sum_{e_j \sim j, e_j \neq e}\frac{z_j}{\sqrt{w_e w_{e_j}}}\right),
    \label{eq:FRC}
\end{equation}
where \( z_i, z_j \) represent node weights and \( w_e \) denotes edge weights corresponding to neural connectivity strength. Positive curvature values typically identify edges in densely connected regions, whereas negative curvature highlights edges that bridge highly connected network modules. 

Ollivier-Ricci curvature (ORC) represents an alternative notion of discrete Ricci curvature~\citep{Ol}, which provides a comparable characterization of network geometry~\citep{samal2018comparative}; we defer a formal definition to Appendix~\ref{apx:ORC}. Its definition via Markov chains lends itself to another interpretation in the context of inter-brain connectivity: The curvature of an edge provides a proxy for its tendency to \textit{attract} information flow, in the sense that negative curvature indicates more attraction~\citep{wang_applying_2022}. Regions with a high density of edges with low (negative) curvature promote shortest-path traversal, while regions with higher (positive) curvature promote diffusion. 

In the next section, we investigate how a toolkit based on discrete Ricci curvature can be fruitfully applied to social neuroscience.

\section{The Case of Hyperscanning}
\label{sec:GeoHyp}

Hyperscanning, defined as the simultaneous recording of neural signals from interacting individuals \citep{Montague2002}, has reshaped social neuropsychology \citep{Schilbach2025} and clinical neuroscience alike \citep{Adel2025}. Despite these advances, the analytical methods applied in hyperscanning remain heavily reliant on purely correlational approaches \citep{Hamilton2021}, inherently restricting their explanatory power. \emph{We contend that the curvature-based analysis of interbrain coupling networks can move hyperscanning studies closer toward mechanistic explanations.} 

\subsection{Interbrain Networks and their Synchrony}
\label{sec:InterNetworks}

Interbrain networks represent the joint neural connectivity of two or more individuals as interconnected nodes within weighted graphs, constructed via hyperscanning, with each node typically corresponding to a neural region and the edge weights derived by computing IBS metrics (e.g., PLV) from the neural activity in these regions \citep{Hakim2023}. These studies have been limited in the mechanistic inferences they afford researchers; at best, correlations between brain regions of interacting subjects can be interpreted in terms of the computational-cognitive roles ascribed to these regions, with detailed mechanisms and their dynamic evolution -- as social behavior unfolds over time -- remaining speculative. We propose extending studies of time-varying interbrain networks with graph curvatures to detect meaningful phase transitions in interpersonal neural dynamics and provide insight into the information routing strategies interbrain networks use to accomplish joint behavioral tasks. We explore these applications in detail in the following sections.

Operationally, a curvature-based hyperscanning analysis proceeds in four steps. First, after standard preprocessing and artifact rejection, interbrain connectivity is estimated in a time-resolved fashion, for example by computing IBS measures in sliding windows and in task-relevant frequency bands. Second, each window is represented as a weighted bipartite or multilayer graph whose nodes correspond to regions in each participant and whose edges reflect interbrain coupling strength. Third, discrete curvatures are computed for edges or nodes in each graph, yielding a time series of curvature distributions that summarizes how interbrain geometry evolves. Finally, summary statistics of these distributions (e.g., entropy, moments, or the proportion of strongly negative edges) are aligned with behavioral events and task structure, and evaluated against appropriate null models (such as trial shuffling or surrogate coupling matrices) to determine whether observed transitions exceed those expected from stationary dynamics or measurement noise.

\subsection{Capturing Phase Transitions}
\label{sec:PhaTrans}

\begin{figure}[t]
\centering
\includegraphics[width=\linewidth]{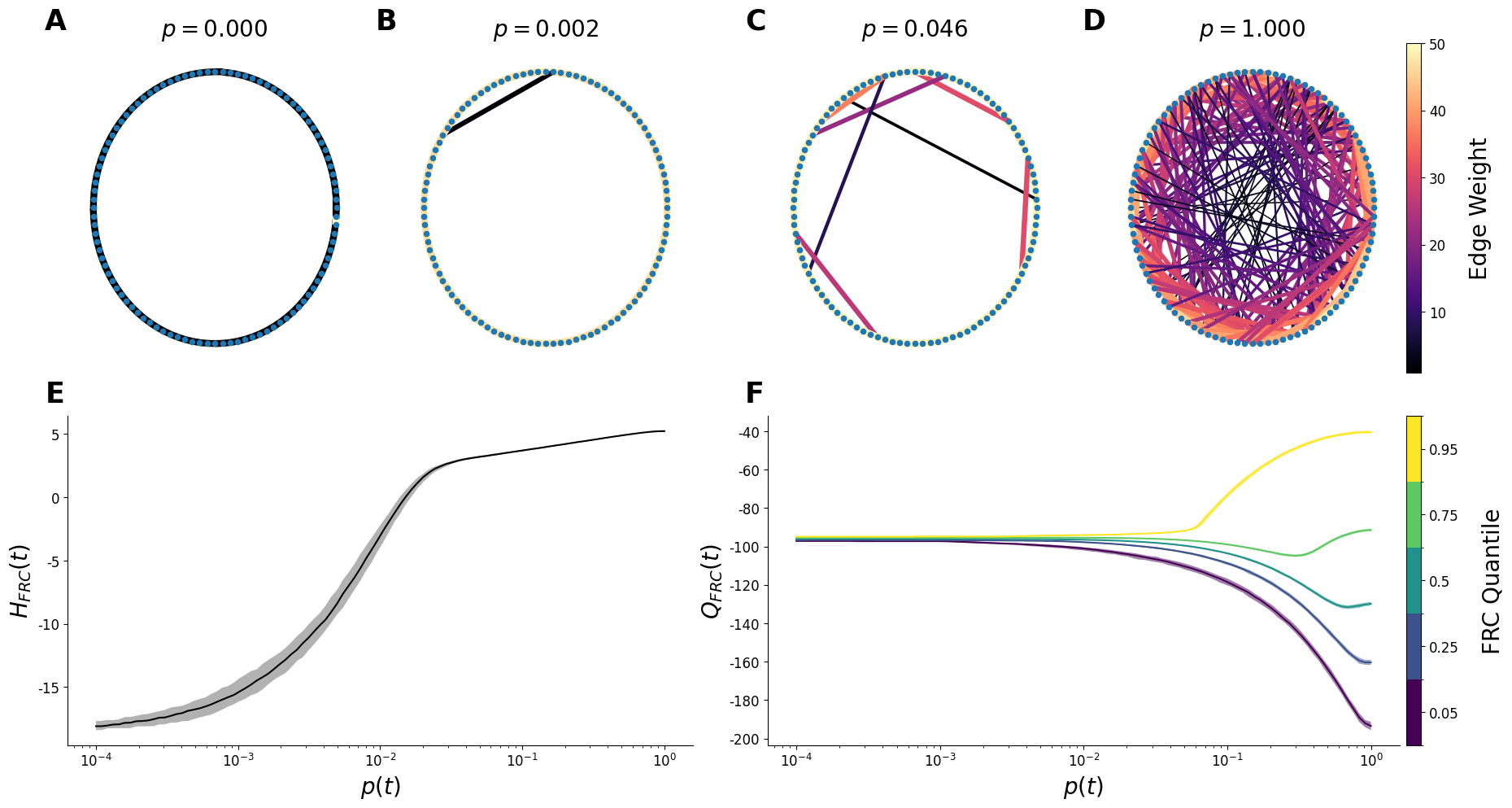}
\caption{Simulations of time-varying brain networks modeled as weighted small-world networks with varying rewiring probability. \textbf{A–D:} Four examples with $N=100$, mean degree $K=5$, and different $p$, generated using \citep{Muldoon2016-mg}. \textbf{E:} Entropy of the FRC distribution as $p$ evolves from 0 to 1 for $N=1000$, $K=50$ (note phase transition around $p=10^{-2}$). \textbf{F:} Corresponding quantiles of the FRC distribution. Solid curves show the median over 200 replications; shaded areas mark 0.05 and 0.95 quantiles.}
\vspace{-2\baselineskip}
\label{fig:SmallWorld}
\end{figure}

Suppose the timing of task-related behavioral transitions or events, such as cooperative engagements, misunderstandings, or conflict resolutions, is synchronized with the timing of phase transitions in interbrain networks as identified by graph curvatures. In that case, investigators can more confidently make inferences about the neural mechanisms of behavior \citep{steyn-ross_modeling_2010}. To capture significant dynamic shifts in network configurations, we examine divergences over time in the differential entropy of graph curvature distributions of IBS, \( H_{RC} \), defined as
\begin{equation}
    H_{RC}(G_t) = -\int_\mathbb{R} f_{RC}^t(x) \log \left[ f_{RC}^t(x) \right] \, dx,
    \label{eq:entropy}
\end{equation}
where \( f_{RC}^t(x) \) describes the probability density of discrete curvature values across the network configuration \( G_t \) at time \( t \)~\citep{znaidi_unified_2023}. In Figure~\ref{fig:SmallWorld}, we apply this method to detect phase transitions in a toy model of time-varying brain networks with small-world topology. We show that as the rewiring probability used to generate the networks evolves from zero to unity, the differential entropy of the FRC distribution undergoes a divergence between $p = 10^{-3}$ and $p = 10^{-1}$ as the network transitions between a regular lattice and a random network. Panels~E–F show a sharp rise in entropy once $p \gtrsim 10^{-2}$ and a widening curvature distribution (95th-percentile jump), due to increased neighborhood overlap and shortcut formation, marking a transition from a segregated, lattice‑like topology to a more integrated small‑world/random regime; see Table~\ref{tab:modalities} in Appendix~\ref{apx:modalities} for modality‑ and condition‑specific expectations that map these geometric signatures to EEG, fNIRS, and fMRI hyperscanning in resting and experimental task conditions.

In empirical hyperscanning data, the rewiring probability $p$ is replaced by latent changes in interpersonal coordination driven by the task. The curvature-entropy trace $H_{RC}(G_t)$ can be inspected for sharp excursions, but in practice these should be quantified by change-point detection, clustering of network states, or model comparison between stationary and non-stationary curvature processes. Critically, the interpretation of a detected ``phase transition'' depends on its alignment with independently measured behavioral markers (onset of joint action, breakdowns in coordination, feedback delivery) and on comparison with subject- and task-specific null ensembles. Without these controls, large curvature fluctuations could simply reflect transient changes in signal-to-noise ratio, motion confounds, or common-input effects rather than genuine reconfigurations of interbrain communication structure.

\subsection{Capturing Information Routing Strategies}
\label{sec:InfRout}

Theoretical work on information routing in brain networks has used Markov chains to model a spectrum of information routing strategies between shortest-path traversal to a target node at one extreme, and random diffusion at the other \citep{Avena2019}. Thus, when applied to interbrain networks, the ORC distribution of the network can be interpreted as identifying the information routing strategy adopted by its subnetworks.

Recent work in deep learning has shown that FRC can identify information bottlenecks that distort information flow during message-passing in graph neural networks \citep{topping,fesser}. These results suggest that FRC could be a valuable tool for assessing information flow in brain networks, a key component of the mechanistic models proposed by predictive theories of the brain \citep{friston_graphical_2017}.

Practically, ORC-based routing analyses can be implemented by tracking, over time, the fraction of edges whose curvature falls below a chosen negative threshold, or by clustering edges and nodes into modules with similar curvature profiles. Intervals dominated by strongly negative curvature subnetworks would then correspond to regimes in which information is funneled along a small number of backbone paths, whereas intervals with predominantly positive curvature would favor more diffuse, exploratory exchange. In a cooperative joint-action task, for instance, one might expect transient emergence of negatively curved ``bridges'' between premotor and parietal regions across brains at moments of tight sensorimotor alignment, followed by relaxation toward more positively curved, segregated configurations during periods of rest or independent planning.

\section{Towards an Interbrain Geometry}

Adopting a geometric framework within neuroscience offers methodological and conceptual advancements over traditional IBS-based analyses. \textit{Geometric hyperscanning} could address the inability of correlation-based metrics alone to capture dynamic network reconfigurations and characterize real-time information routing strategies within and between socially interacting brains. 

Discrete curvature distributions could summarize constraints on network dynamics, with divergences in the entropy of the distribution indicating network reorganization events. While this does not intrinsically resolve the confounding factors that arise in IBS-based approaches, it provides a complementary network-level description of interbrain interactions, enabling further inferences required to construct mechanistic explanations. This direction accords with \citet{Kulkarni2024}’s call for minimal, principled models of brain-network complexity and with \citet{Sporns2010}’s emphasis on meso-scale features (hubs, clusters, bridges), reinforcing discrete curvatures as indicators of structural transitions during social interaction. 

Several limitations and open questions remain. Curvature estimates are sensitive to how interbrain graphs are constructed, including choices of frequency band, time-window length, edge-weight transformation, and thresholding; systematic benchmarking across these design decisions is still lacking. Different notions of discrete curvature may capture partially non-overlapping aspects of network organization, raising the question of whether a single curvature measure suffices or whether a multi-geometry description is required. 

Curvature-based analyses could allow researchers to explore the information routing implications of IBS and how they reorganize dynamically throughout real-time interactions, as captured in hyperscanning data, paving the way for a deeper mechanistic understanding of the social brain.

\bibliography{gtml2025_workshop}

@article{Adel2025,
  title   = "A systematic review of hyperscanning in clinical encounters",
  author  = "Lena Adel and Lisane Moses and Elisabeth Irvine and Kyle T. Greenway and Guillaume Dumas and Michael Lifshitz",
  journal = "Neuroscience \& Biobehavioral Reviews",
  volume  = "176",
  pages   = "106248",
  year    = 2025,
  doi     = "10.1016/j.neubiorev.2025.106248",
  url     = "https://doi.org/10.1016/j.neubiorev.2025.106248"
}

@article{Forman2003,
  title   = "Bochner's Method for Cell Complexes and Combinatorial Ricci Curvature",
  author  = "R. Forman",
  journal = "Discrete and Computational Geometry",
  volume  = "29",
  pages   = "323-374",
  year    = 2003,
  doi     = "https://doi.org/10.1007/s00454-002-0743-x",
  url     = "https://doi.org/10.1007/s00454-002-0743-x"
}

@article{Hakim2023,
  title   = "Quantification of inter-brain coupling: A review of current methods used in haemodynamic and electrophysiological hyperscanning studies",
  author  = "U. Hakim and S. De Felice and P. Pinti and X. Zhang and J. A. Noah and Y. Ono and P. W. Burgess and A. Hamilton and J. Hirsch and I. Tachtsidis",
  journal = "NeuroImage",
  volume  = "280",
  pages   = "120354",
  year    = 2023,
  doi     = "10.1016/j.neuroimage.2023.120354",
  url     = "https://doi.org/10.1016/j.neuroimage.2023.120354"
}

@article{Hamilton2021,
  title   = "Hyperscanning: Beyond the Hype",
  author  = "A. F. C. Hamilton",
  journal = "Neuron",
  volume  = "109",
  number  = "3",
  pages   = "404-407",
  year    = 2021,
  doi     = "10.1016/j.neuron.2020.11.008",
  url     = "https://doi.org/10.1016/j.neuron.2020.11.008"
}

@article{Hinrichs2025,
  title   = "Geometric Hyperscanning of Affect under Active Inference",
  author  = "Nicolás Hinrichs and Mahault Albarracin and Dimitris Bolis and Yuyue Jiang and Leonardo Christov-Moore and Leonhard Schilbach",
  journal = "arXiv preprint arXiv:2506.08599",
  year    = 2025,
  url     = "https://arxiv.org/abs/2506.08599"
}

@article{Kulkarni2024,
  title={Towards principles of brain network organization and function},
  author={Kulkarni, Suman and Bassett, Dani S},
  journal={arXiv preprint arXiv:2408.02640},
  year={2024},
  url={https://arxiv.org/abs/2408.02640}
}

@article{Montague2002,
  title   = "Hyperscanning: Simultaneous fMRI during Linked Social Interactions",
  author  = "P. Read Montague and Gregory S. Berns and Jonathan D. Cohen and Samuel M. McClure and Giuseppe Pagnoni and Mukesh Dhamala and Michael C. Wiest and Igor Karpov and Richard D. King and Nathan Apple and Ronald E. Fisher",
  journal = "NeuroImage",
  volume  = "16",
  number  = "4",
  pages   = "1159-1164",
  year    = 2002,
  doi     = "10.1006/nimg.2002.1150",
  url     = "https://doi.org/10.1006/nimg.2002.1150"
}

@article{Schilbach2025,
  title   = "Synchrony Across Brains",
  author  = "Leonhard Schilbach and Elizabeth Redcay",
  journal = "Annual Review of Psychology",
  volume  = "76",
  pages   = "883-911",
  year    = 2025,
  doi     = "10.1146/annurev-psych-080123-101149",
  url     = "https://doi.org/10.1146/annurev-psych-080123-101149"
}

@book{Sporns2010,
  title={Networks of the Brain},
  author={Sporns, Olaf},
  year={2010},
  publisher={MIT Press},
  address={Cambridge, MA}
}

@article{Weber2019,
      title={Curvature-based Methods for Brain Network Analysis}, 
      author={Melanie Weber and Johannes Stelzer and Emil Saucan and Alexander Naitsat and Gabriele Lohmann and Jürgen Jost},
      year={2019},
      eprint={1707.00180},
      journal={arXiv},
      primaryClass={q-bio.NC},
      url={https://arxiv.org/abs/1707.00180}, 
}

@article{Weber2025,
author = {Weber, Melanie},
title = {Geometric Machine Learning},
journal = {AI Magazine},
volume = {46},
number = {1},
pages = {e12210},
doi = {https://doi.org/10.1002/aaai.12210},
url = {https://onlinelibrary.wiley.com/doi/abs/10.1002/aaai.12210},
eprint = {https://onlinelibrary.wiley.com/doi/pdf/10.1002/aaai.12210},
year = {2025}
}

@article{Avena2019, 
    title = "A spectrum of routing strategies for Brain Networks", 
    url = "https://journals.plos.org/ploscompbiol/article?id=10.1371%2Fjournal.pcbi.1006833", 
    journal = "PLOS Computational Biology",
    author = {Avena-Koenigsberger, Andrea and Yan, Xiaoran and Kolchinsky, Artemy and Heuvel, Martijn P. van den and Hagmann, Patric and Sporns, Olaf}, 
    year = {2019}, 
    month = {Mar}
}

@article{Muldoon2016-mg,
  title     = "{Small-World} Propensity and weighted brain networks",
  author    = "Muldoon, Sarah Feldt and Bridgeford, Eric W and Bassett,
               Danielle S",
  journal   = "Sci. Rep.",
  volume    =  6,
  number    =  1,
  pages     = "22057",
  month     =  {Feb},
  year      =  {2016},
}

@article{znaidi_unified_2023,
	title = {A unified approach of detecting phase transition in time-varying complex networks},
	volume = {13},
	issn = {2045-2322},
	url = {https://www.nature.com/articles/s41598-023-44791-3},
	doi = {10.1038/s41598-023-44791-3},
	number = {1},
	journal = {Scientific Reports},
	author = {Znaidi, Mohamed Ridha and Sia, Jayson and Ronquist, Scott and Rajapakse, Indika and Jonckheere, Edmond and Bogdan, Paul},
	month = oct,
	year = {2023},
	pages = {17948}
}

@book{steyn-ross_modeling_2010,
	address = {New York, NY},
	title = {Modeling Phase Transitions in the Brain},
	isbn = {978-1-4419-0795-0 978-1-4419-0796-7},
	publisher = {Springer},
	editor = {Steyn-Ross, D. Alistair and Steyn-Ross, Moira},
	year = {2010},
	doi = {10.1007/978-1-4419-0796-7}
}

@article{wang_applying_2022,
	title = {Applying Ollivier-Ricci curvature to indicate the mismatch of travel demand and supply in urban transit network},
	volume = {106},
	issn = {1569-8432},
	url = {https://www.sciencedirect.com/science/article/pii/S0303243421003731},
	doi = {10.1016/j.jag.2021.102666},
	journal = {International Journal of Applied Earth Observation and Geoinformation},
	author = {Wang, Yaoli and Huang, Zhou and Yin, Ganmin and Li, Haifeng and Yang, Liu and Su, Yuelong and Liu, Yu and Shan, Xv},
	month = {Feb},
	year = {2022},
	pages = {102666},
}

@article{friston_graphical_2017,
	title = {The graphical brain: Belief propagation and active inference},
	volume = {1},
	issn = {2472-1751},
	url = {https://www.ncbi.nlm.nih.gov/pmc/articles/PMC5798592/},
	doi = {10.1162/NETN_a_00018},
	number = {4},
	journal = {Network Neuroscience},
	author = {Friston, Karl J. and Parr, Thomas and de Vries, Bert},
	month = {Dec},
	year = {2017},
	pages = {381--414},
}

@article{samal2018comparative,
  title={Comparative analysis of two discretizations of Ricci curvature for complex networks},
  author={Samal, Areejit and Sreejith, RP and Gu, Jiao and Liu, Shiping and Saucan, Emil and Jost, J{\"u}rgen},
  journal={Scientific reports},
  volume={8},
  number={1},
  pages={8650},
  year={2018},
  publisher={Nature Publishing Group UK London}
}

@article{Ol,
  title={Ricci curvature of Markov chains on metric spaces},
  author={Ollivier, Y.},
  journal={Journal of Functional Analysis},
  volume={256},
  number={3},
  pages={810--864},
  year={2009},
  publisher={Elsevier}
}

@inproceedings{topping,
	title = {Understanding over-squashing and bottlenecks on graphs via curvature},
	booktitle={International Conference on Learning Representations},
	author = {Topping, Jake and Giovanni, Francesco Di and Chamberlain, Benjamin Paul and Dong, Xiaowen and Bronstein, Michael M.},
	year = {2022}
}

@inproceedings{fesser,
  title={Mitigating over-smoothing and over-squashing using augmentations of Forman-Ricci curvature},
  author={Fesser, Lukas and Weber, Melanie},
  booktitle={Learning on Graphs Conference},
  year={2023}
}

@article{chatterjee2021detecting,
  title={Detecting network anomalies using Forman--Ricci curvature and a case study for human brain networks},
  author={Chatterjee, Tanima and Albert, R{\'e}ka and Thapliyal, Stuti and Azarhooshang, Nazanin and DasGupta, Bhaskar},
  journal={Scientific reports},
  volume={11},
  number={1},
  pages={8121},
  year={2021},
  publisher={Nature Publishing Group UK London}
}

\clearpage     
\appendix

\section{Ollivier's Ricci Curvature}
\label{apx:ORC}

We provide a formal definition of Ollivier's Ricci curvature, which was discussed in the main text. 

Consider the 1-hop neighborhoods of two adjacent nodes $u$ and $v$ in a network and equip each with uniform measures defined as follows: Let $m_{u}(i) := \frac{z_i}{\sum_{j \in \mathcal{N}_u} z_j}$, where $i$ is a neighbor of $u$, $z_i$ its weight, and $\mathcal{N}_u$ denotes $u$'s 1-hop neighborhood. An analogous measure can be defined on the neighborhood of $v$. The cost of transporting mass between these two node neighborhoods along the edge $e=(u,v)$ is quantified by the Wasserstein-1 distance between the measures, namely
\begin{equation}
    W_1(m_{u}, m_{v}) = \inf_{m \in \Gamma(m_{u},m_{v})} \int_{(z,z') \in V \times V} d(z,z') \, m(z,z') \, dz \, dz' \, ,
\end{equation}
where $\Gamma(m_{u},m_{v})$ is the set of all measures over $V \times V$ whose marginals are $m_{u}$ and $m_{v}$.  
The \emph{Ollivier-Ricci curvature}~\citep{Ol} is then defined as
\begin{equation}\label{eq:orc-e}
	\kappa(u,v) := 1 - \frac{W_1(m_{u}, m_{v})}{d_G(u,v)} \, ,
\end{equation}
with $d_G(u,v)$ denoting the shortest-path distance between $u$ and $v$ in $G$.

\section{Hyperscanning Modalities Across Conditions}
\label{apx:modalities}

We provide a comparative overview pairing illustrative values to empirical hyperscanning modalities across common conditions, as drawn from our simulations.  

\renewcommand{\arraystretch}{1.1}
\begin{table}[h]
\small
\setlength{\tabcolsep}{4pt}
\centering
\begin{tabular}{lp{0.18\linewidth}p{0.55\linewidth}}
\toprule
\textbf{Mod./Cond.} & \textbf{Edge-Weight Range} & \textbf{Empiric Implication} \\
\midrule
EEG -- Task     & PLV $\approx 0.2$--$0.6$ & fast, captures rapid behavior \\
EEG -- Resting  & PLV $\approx 0.1$--$0.4$ & fast, spontaneous activity \\
fNIRS -- Task   & Corr. $\approx 0.1$--$0.3$ & 0.1--1 s, suited to slow tasks \\
fNIRS -- Resting& Corr. $<0.2$ & slow, long-term fluctuations \\
fMRI -- Task    & Cohe. $\approx 0.2$--$0.5$ & 1--2 s, block tasks only (too slow for fast actions) \\
fMRI -- Resting & Cohe. $<0.2$ & very slow, long-term networks \\
\bottomrule
\end{tabular}
\caption{Modalities across conditions, their canonical edge-weight ranges, and characteristic spatiotemporal scales with empirical implications.}
\label{tab:modalities}
\end{table}

The modality-dependent spatiotemporal sampling rates and signal strengths frame the core challenge addressed by our pipeline: detecting network reconfigurations only when the neural signal is sampled at a sufficient rate to resolve the target behavior.

\end{document}